\documentclass[aps,pra,twocolumn,amsfonts,amssymb,amsmath,showpacs,
floatfix,nofootinbib,groupedaddress,superscriptaddress,citesort]{revtex4}
\usepackage{mathrsfs}
\usepackage{amsfonts}
\usepackage{amstext}
\usepackage{amsmath}
\usepackage{amssymb}
\usepackage[usenames]{xcolor}
\usepackage[dvips]{graphicx}
\def\qed{\leavevmode\unskip\penalty9999 \hbox{}\nobreak\hfill
     \quad\hbox{\leavevmode  \hbox to.77778em{%
              \hfil\vrule   \vbox to.675em%
               {\hrule width.6em\vfil\hrule}\vrule\hfil}}
     \par\vskip3pt}

\begin{document}
\title{Assisted state  discrimination without entanglement}

\author{Bo Li}
\email{libo@iphy.ac.cn}
\affiliation{Institute of Physics, Chinese Academy of Sciences, Beijing 100190, China}
\affiliation{Department of Mathematics and Computer, Shangrao Normal University,
 Shangrao 334001, China}
 \author{Shao-Ming Fei}
\affiliation{School of Mathematical Sciences, Capital Normal
University, Beijing 100048, China} \affiliation{Max-Planck-Institute
for Mathematics in the Sciences, 04103 Leipzig, Germany}
\author{Zhi-Xi Wang}
\affiliation{School of Mathematical Sciences, Capital Normal
University, Beijing 100048, China}
\author{Heng Fan}
\email{hfan@iphy.ac.cn}
\affiliation{Institute of Physics, Chinese Academy of Sciences, Beijing 100190, China}

\begin{abstract}
It is shown that the dissonance,
a quantum correlation which is equal to quantum discord for separable state, is required for
assisted optimal state discrimination.
%Roa \emph{et.al}[Phys. Rev. Lett. 107, 080401(2011)] find that
We find that only one side discord is required in  the optimal process of assisted state discrimination,
while another side discord and entanglement is not necessary.
We confirm that the quantum discord, which is asymmetric depending on local measurements,
is a resource for assisted state discrimination.
With the absence of entanglement, we give the necessary and sufficient condition
for vanishing one side discord in assisted state discrimination
for a class of $d$ nonorthogonal states.
As a byproduct, we find that the positive-partial-transposition (PPT)
condition is the necessary and sufficient condition for the separability of a class of $2\times d$ states.

\end{abstract}
\maketitle
\section{Introduction}

Entanglement is regarded as a key resource in quantum information processing such as
teleportation and superdense coding, etc.\cite{Nielsen&Chuang}.
On the other hand,
it is shown that a deterministic quantum computation
with one qubit(DQC1) \cite{Knill} can be carried out without entanglement.
While the quantum discord \cite{Zurek,Vedral,horodecki,Luo1,Caves2},
another type of quantum correlation, might be the reason
for the advantage of this algorithm that surpasses the corresponding classical algorithms.
Besides quantum entanglement, there are also many quantum nonlocal properties which can be manifested by
separable states or separable operations \cite{Bennett-I-99}.
It is thus reasonable to assume that the quantumness correlation can be viewed
from different aspects.
Recently much efforts have been devoting to studying
various measures of nonclassical correlation, see for example Refs.\cite{SKB,PWinter,SKBnew,giorgi,gessner,Rossatto,Luo84,tao,Jian}.
Quantum discord, which is our main concern in this paper,
is recently pointed out to have operational
interpretations, for example in terms of the quantum state merging protocol~\cite{Cavalcanti}
and as entanglement by an activation protocol or by measurement\cite{PWinter,SKB}.

Quantum discord \cite{Vedral,Zurek}
is measured by the difference between the mutual information and the
maximal conditional mutual information obtained by local measurement.
Explicitly, let us consider a bipartite quantum state $\rho_{AB}$,  the ``right'' quantum discord is given by~\cite{Vedral,Zurek},
\begin{eqnarray}\label{rightdis}
D_B(\rho_{AB})=I(\rho_{AB})-\sup_{E_k}\{S(\rho_A)-\sum_k p_{A|k} S(\rho_{A|k})\},
\end{eqnarray}
where $S(\rho)$ is the von Neumann entropy and  $I(\rho_{AB})=S(\rho_A)+S(\rho_B)-S(\rho_{AB})$ is the quantum mutual information,
and $p_{A|k}={\rm Tr}(\openone_A\otimes E_k\rho)$,  $\rho_{A|k}={\rm Tr}_B(\openone_A\otimes E_k\rho)/p_{A|k}$, the supreme is taken
over all the von Neumann  measurement sets $\{E_k\}$ on system $B$.
Similarly, $D_A(\rho_{AB})$
refers to the ``left'' discord and is given by
\begin{eqnarray}\label{leftdis}
D_A(\rho_{AB})=I(\rho_{AB})-\sup_{E_k}\{S(\rho_B)-\sum_k p_{B|k} S(\rho_{B|k})\},
\end{eqnarray}
with $p_{B|k}={\rm Tr}(E_k\otimes\openone_B\rho)$,  $\rho_{B|k}={\rm Tr}_A((E_k\otimes\openone_B\rho)/p_{B|k}$.
Note that the difference between those two discords is that the measurement is performed
on party $A$ or on party $B$, respectively. It is thus expected that
this definition of quantum discord is
not symmetric with respect to $A$ and $B$.

Due to the supreme in the definition (\ref{rightdis}),  quantum discord does not have  analytic
or operational expression in general.  However, explicit expression of discord
has been found for cases like Bell-diagonal state~\cite{Luo,lang} and subclass
of so-called $X$ states~\cite{Ali,bo,qingchen}. Numerical results for general two-qubit states are also presented in \cite{girolami}.
Moreover, the quantum discord can be related to
the entanglement of formation (EOF) \cite{Koashi}. This relation can be used to calculate the quantum discord for
 the rank-two states \cite{shi}. Nevertheless, zero discord
 can be easily found by a necessary and sufficient condition pointed out in Ref.\cite{dakic}.
 If $\rho_{AB}$ is a quantum state of $d_{A}\times d_{B}$ dimensional bipartite quantum systems, from \cite{dakic}
 $\rho_{AB}$ can be written in diagonal form $\rho_{AB}=\sum_{n=1}^Lc_nS_n\otimes F_n$,
 where $\{S_n\},\{F_m\}, (n=1,\cdots,d_{A}^2,m=1,\cdots,d_{B}^2)$ are the bases of the respective
 local spaces, $L$ is the rank of $\rho$\cite{dakic}. Then the necessary and sufficient
 condition of zero ``left'' discord is
  \begin{eqnarray}\label{zeorleftdis}
[S_i,S_j]=0, i,j=1,\cdots,L.
\end{eqnarray}
Similarly, the necessary and sufficient condition of zero ``right'' discord is
\begin{eqnarray}\label{zeorrightdis}
[F_i,F_j]=0, i,j=1,\cdots,L.
\end{eqnarray}
 It is shown that almost all quantum states have nonzero discord~\cite{ferraro,yuchun}.

A unified view of quantum correlation based on relative entropy was introduced in~\cite{kavanmodi}.
 Quantum dissonance is a kind of quantum correlation of separable states. As the closest state to a
 separable state $\rho$ as measured by relative entropy is the state $\rho$
itself, the dissonance is just the quantum discord in this case.

It is remarkable that dissonance is found to be useful in assisted state discrimination recently \cite{Roa}. Note that dissonance is the quantum discord in the optimal case.
In this paper, we further find that actually only one side quantum discord appears in the optimal process,
while discord of another side can be zero. Hence the role of quantum discord played in
assisted state discrimination might be quite
different for different parties. We show clearly that the quantum discord is really
a useful resource which can be used in assisted state discrimination.
We also extend the study of discord for assisted state discrimination
to more general cases.

The paper is organized as follows.  In Sec.~\ref{surface},
we first give a brief review of the model in assisted state discrimination. We then
show that only one side discord is necessary in the optimal case.
In Sec.~\ref{dynamics} and Sec.~\ref{optimal}, we generalize the model to discriminate a class of
$d$ nonorthogonal states, in which we still find that one-side discord is required in the process.
Sec.~\ref{discuss} is the summary.

\section{Unambiguous discrimination of  two nonorthogonal states  }\label{surface}
Following the model of assisted state
discrimination in Ref.~\cite{peres,jaeger,Roa},  consider that a qubit is randomly prepared in one
of the two nonorthogonal states $|\psi_+\rangle$ or $|\psi_-\rangle$ with a \emph{priori} probabilities
$p_+$ and $p_-$ with $p_++p_-=1$. Our aim is to discriminate the two states $|\psi_+\rangle$ or $|\psi_-\rangle$.
The system is coupled to an auxiliary qubit $A$ by a joint unitary
transformation $U$ such that
\begin{eqnarray}
U|\psi_+\rangle|k\rangle_a=\sqrt{1-|\alpha_+|^2}|+\rangle|0\rangle_a+\alpha_+|0\rangle|1\rangle_a, \nonumber\\
U|\psi_-\rangle|k\rangle_a=\sqrt{1-|\alpha_-|^2}|-\rangle|0\rangle_a+\alpha_-|0\rangle|1\rangle_a,
\label{unitary}
\end{eqnarray}
where $|k\rangle_a$ is an auxiliary state with orthonormal basis $\{|0\rangle_a,|1\rangle_a\}$,
$|\pm\rangle\equiv (|0\rangle\pm|1\rangle)/\sqrt{2}$ are the orthonormal states of the system
that can be discriminated. For convenience, we adopt the symbols used in Ref.~\cite{Roa}.
A \emph{priori} fixed overlap is, $\langle\psi_+|\psi_-\rangle=\alpha=|\alpha|e^{i\theta}=\alpha_+^\ast\alpha_-$,
where $\alpha_+^\ast$ is the complex conjugate of $\alpha_+$, $\theta$ is the phases of $\alpha$. The mixed state
we consider in discrimination is given by
\begin{eqnarray}\label{rho}
\rho_{|\alpha_+|} & = & p_+U\left( |\psi_+\rangle\langle\psi_+|\otimes |k\rangle_a \langle k|\right) U^\dag\nonumber\\
&  & + p_-U\left( |\psi_-\rangle \langle\psi_-|\otimes |k\rangle_a \langle k|\right) U^\dag.
\end{eqnarray}
By performing a von Neumann measurement on the
auxiliary system by basis, $\{|0\rangle_a\langle0|,|1\rangle_a\langle1|\}$, the auxiliary state in (\ref{rho}) will
collapse to either $\{|0\rangle_a\langle0|\}$ or $\{|1\rangle_a\langle1|\}$.
In case the system collapses to $\{|0\rangle_a\langle0|\}$,
we will discriminate successfully the original state since we can distinguish
deterministically the states $|\pm \rangle $ as in (\ref{unitary}). The success probability is given by
\begin{eqnarray}\label{prob}
P(|\alpha_+|)=1-p_-\frac{|\alpha|^2}{|\alpha_+|^2}-p_+|\alpha_+|^2.
\end{eqnarray}
For general $|\alpha_+|$, the probability is neither 0 nor 1.

We shall use the PPT criterion ~\cite{peres1}
to characterize  the separability of $\rho_{|\alpha_+|}$.
For $2\times 2$ and $2\times 3$ systems,  the PPT condition
is both necessary and sufficient for the separability of quantum states \cite{Zyczkowski}.
From the PPT criterion,
$\rho_{|\alpha_+|}$ is separable if and only if its partial transposed matrix $\rho_{|\alpha_+|}^{T_A}$ has
a non-negative spectrum. Hence the determinant $D$ of $\rho_{|\alpha_+|}^{T_A}$ should be non-negative either.
By direct calculation we have
 \begin{eqnarray*}
 D& &=-(p_+\frac{1-|\alpha_+|^2}{2}+p_-\frac{1-|\alpha_-|^2}{2})\times(p_+|\alpha_+|^2+ \nonumber\\
  & & p_-|\alpha_-|^2)\times |p_+\sqrt{\frac{1-|\alpha_+|^2}{2}}\alpha_+-p_-\sqrt{\frac{1-|\alpha_-|^2}{2}}\alpha_-|^2 .
\end{eqnarray*}
$D\geq 0$ implies that
\begin{eqnarray}\label{sepa}
p_+\sqrt{\frac{1-|\alpha_+|^2}{2}}\alpha_+=p_-\sqrt{\frac{1-|\alpha_-|^2}{2}}\alpha_-,
\end{eqnarray}
that is,
\begin{eqnarray}\label{sepa1}
p_+\sqrt{\frac{1-|\alpha_+|^2}{2}}|\alpha_+|^2=p_-\sqrt{\frac{1-|\alpha_-|^2}{2}}\alpha,
\end{eqnarray}
where $\alpha=\alpha_+^\ast\alpha_-$. From (\ref{sepa1}) we know that $\alpha$ must be a real number.
Eq.(\ref{sepa1}) is a necessary condition for the separability of $\rho_{|\alpha_+|}$
and is actually also a sufficient condition.
It is in fact the same as the Eq.(7) in ~\cite{Roa}.

That the condition (\ref{sepa}) is also a sufficient condition for separability can be seen from the
following separable form of $\rho_{|\alpha_+|}$,
\begin{eqnarray}\label{rhosepa}
\rho_{|\alpha_+|} &=&(1-p_+|\alpha_+|^2-p_-|\alpha_-|^2)\rho_1^S\otimes |0\rangle_a\langle0|+\nonumber\\
& &  (p_+|\alpha_+|^2+p_-|\alpha_-|^2)|0\rangle\langle0|\otimes \rho_2^A,
\label{exform}
\end{eqnarray}
where $\rho_1^S$ and $\rho_2^A$ are the density matrices of the principal system and the auxiliary system
respectively,
\begin{eqnarray*}
\rho_1^S & = & \frac{1}{1-p_+|\alpha_+|^2-p_-|\alpha_-|^2}(p_+(1-|\alpha_+|^2)|+\rangle\langle+|\nonumber\\
  &  & +p_-(1-|\alpha_-|^2)|-\rangle\langle-|),\nonumber\\
\rho_2^A & = & \frac{1}{p_+|\alpha_+|^2+p_-|\alpha_-|^2}((p_+|\alpha_+|^2+p_-|\alpha_-|^2)|1\rangle_a\langle1|\nonumber\\
 & & +\sqrt{2}p_+\alpha_+\sqrt{1-|\alpha_+|^2}|1\rangle_a\langle0|+\nonumber\\
 &  & \sqrt{2}p_+\alpha_+^\ast\sqrt{1-|\alpha_+|^2}|0\rangle_a\langle1|).
\end{eqnarray*}
From (\ref{exform}) and the necessary and sufficient condition of zero discord given by Eq.(\ref{zeorleftdis}) and (\ref{zeorrightdis}), we have that
state $\rho_{|\alpha_+|}$ has
zero ``right'' quantum discord when the components of the ``right'' reduced density operators are commuting,
$[\rho_2^A ,|0\rangle_a\langle0|]=0$.
This can be satisfied only when $\alpha_+=0$ or $|\alpha_+|=1$.
Since $|\psi_+\rangle$ and $|\psi_-\rangle$ are different nonorthogonal states, we do
not need to consider those two cases because  they corresponds to either the same state or two orthogonal states.
Hence as pointed out in Ref.\cite{Roa}, the ``right'' discord is always non-zero.
On the other hand, $\rho_{|\alpha_+|}$ has
zero ``left'' quantum discord if and only if $[\rho_1^S ,|0\rangle\langle0|]=0$, that is,
\begin{eqnarray}\label{commu}
p_+(1-|\alpha_+|^2)=p_-(1-|\alpha_-|^2).
\end{eqnarray}
Combining (\ref{sepa1}) and (\ref{commu}), we have
\begin{itemize}
\item   $\alpha$ is a real number, and $\alpha\geq 0$;
\item   $p_+=p_-=\frac{1}{2}$;
\item  $|\alpha_+|=|\alpha_-|=\sqrt{|\alpha|}=\sqrt{\alpha}$.
\end{itemize}
It is interesting that those three conditions coincide exactly with the optimal assisted state discrimination case in ~\cite{Roa}
with a \emph{priori} probabilities and $\theta=0$, in which it is shown that
the dissonance is required in the discrimination.
Thus, we obtain one of our main results: The assisted state discrimination of two nonorthogonal states can
be performed with the absence of entanglement. The ``right'' quantum discord is required
for assisted state discrimination. However, the ``left'' discord is not necessarily to be non-zero.
In particular, in the assisted \emph{optimal} state discrimination, the ``left'' discord
is found to be zero. Here we remark that, it is not surprising that the ``left'' discord is
nonzero except for the optimal case, since quantum discord in general is non-zero \cite{ferraro,SKBnew}.

Recall that the motivation of quantum discord is to
find the difference between  total correlations,
including both quantum and classical correlation quantified by mutual information,
and the accessible classical correlation, which is quantified by the maximal
conditional entropy obtained by local measurement \cite{Vedral}. While the mutual
information is symmetric, the asymmetry of the quantum discord is due to local measurements.
The protocol of assisted state discrimination in this paper and in Ref.\cite{Roa}, is
exactly assisted by local measurements on the ``right'' party so as
to distinguish nonorthogonal states of the ``left'' party. Thus the corresponding ``right''
quantum discord is necessary while the ``left'' discord, which is quantified by a local
measurement on the ``left'' party, is useless. Thus we can find that in the optimal assisted
state discrimination, the ``left'' discord is zero. Indeed, the deep
reason that quantum discord is required for assisted state discrimination
is that it is really used, as a consuming resource
in such quantum information processing.

\section{Unambiguous discrimination of  $d$ nonorthogonal states}\label{dynamics}
In the following, we generalize the previous model to $d$-dimensional system.
Let us consider  that a qudit is
randomly prepared in the $d$ ($d\geq2$) nonorthogonal and linearly independent
states, $|\psi_1\rangle, |\psi_2\rangle,\cdots, |\psi_d\rangle$, with a \emph{priori}
probabilities $p_1,p_2, \cdots ,p_d$ with $p_1+\cdots+p_d=1$.
The system is coupled to an auxiliary qubit $A$ by a joint unitary
transformation $U_1$ such that
\begin{eqnarray}
U_1|\psi_1\rangle|k\rangle_a=\sqrt{1-|\alpha_1|^2}|1\rangle|0\rangle_a+\alpha_1\frac{|1\rangle+\cdots+|d\rangle}{\sqrt{d}}|1\rangle_a, \nonumber\\
U_1|\psi_2\rangle|k\rangle_a=\sqrt{1-|\alpha_2|^2}|2\rangle|0\rangle_a+\alpha_2\frac{|1\rangle+\cdots+|d\rangle}{\sqrt{d}}|1\rangle_a, \nonumber\\
\cdots\cdots\cdots\cdots\cdots\cdots\cdots\cdots\cdots\cdots\cdots\cdots\cdots\cdots\cdots\cdots\cdots, \nonumber\\
U_1|\psi_d\rangle|k\rangle_a=\sqrt{1-|\alpha_d|^2}|d\rangle|0\rangle_a+\alpha_d\frac{|1\rangle+\cdots+|d\rangle}{\sqrt{d}}|1\rangle_a,
\label{dunitary}
\end{eqnarray}
where $\{|1\rangle,|2\rangle,\cdots,|d\rangle\}$ is the orthogonal basis in the state space.
Note that the auxiliary state is still two-dimensional.
For $d=2$, we can apply a Hadamard gate on the first qubit,
and the model returns to the one given in Section \ref{surface}.
This model is a natural generalization of discrimination of two nonorthogonal states.
To ensure that there exists the unitary transformation $U_1$,
the inner product of vectors on  the right hand side should be equal to the overlap
of the corresponding original states\cite{duan}, that is
 $\alpha_i^*\alpha_j=\langle\psi_i|\psi_j\rangle$.
 For convenience, we  denote  $\alpha_{ij}=\langle\psi_i|\psi_j\rangle$.

The mixed state we consider in discrimination is given now by
\begin{eqnarray}\label{rhod}
\rho &=&  p_1U_1\left( |\psi_1\rangle\langle\psi_1|\otimes |k\rangle_a\langle k|\right) U_1^\dag
\nonumber \\
&&+ p_2U_1\left( |\psi_2\rangle\langle\psi_2|\otimes |k\rangle_a\langle k|\right) U_1^\dag\nonumber\\
&& +\cdots+p_dU_1\left( |\psi_d\rangle\langle\psi_d|\otimes |k\rangle_a\langle k|\right) U_1^\dag.
\end{eqnarray}
The success probability to discriminate the state is given by
\begin{eqnarray}\label{proo}
P=1-p_1|\alpha_1|^2-p_2|\alpha_2|^2-\cdots-p_d|\alpha_d|^2.
\end{eqnarray}
In general, it is not zero. If $\rho$ is a separable state, the partial transposed matrix must be positive. Then
all the principal minor determinants of $\rho^{T_S}$ is non-negative, where $\rho^{T_S}$ is the
partial transposed matrix with respect to the qudit system. In the following, as a necessary condition
for separability, we calculate those $4\times4$ principal minor determinants that
should be non-negative. Let $M_{ij}$ be the principal minor matrix of $\rho^{T_S}$ by selecting the
$\langle i|\otimes_a\langle 0|,\langle i|\otimes_a\langle 1|,\langle j|\otimes_a\langle 0|,\langle j|\otimes_a\langle 1|$
rows, and the  $|i\rangle\otimes|0\rangle_a,|i\rangle\otimes|1\rangle_a,|j\rangle\otimes|0\rangle_a,|j\rangle\otimes|1\rangle_a$  columns.
By straightforward calculations we have
\begin{widetext}
\begin{eqnarray}
M_{ij} =  \left(
\begin{array}{cccc}
p_i(1-|\alpha_i|^2)
& p_i\alpha_i^*\sqrt{\frac{1-|\alpha_i|^2}{d}} & 0 & p_j\alpha_j^*\sqrt{\frac{1-|\alpha_j|^2}{d}} \\
p_i\alpha_i\sqrt{\frac{1-|\alpha_i|^2}{d}} & \frac{1}{d}(p_1|\alpha_1|^2+\cdots+p_d|\alpha_d|^2) &
p_i\alpha_i\sqrt{\frac{1-|\alpha_i|^2}{d}} & \frac{1}{d}(p_1|\alpha_1|^2+\cdots+p_d|\alpha_d|^2) \\
0 & p_i\alpha_i^*\sqrt{\frac{1-|\alpha_i|^2}{d}} & p_j(1-|\alpha_j|^2)
& p_j\alpha_j^*\sqrt{\frac{1-|\alpha_j|^2}{d}} \\
p_j\alpha_j\sqrt{\frac{1-|\alpha_j|^2}{d}} & \frac{1}{d}(p_1|\alpha_1|^2+\cdots+p_d|\alpha_d|^2) &
 p_j\alpha_j\sqrt{\frac{1-|\alpha_j|^2}{d}} & \frac{1}{d}(p_1|\alpha_1|^2+\cdots+p_d|\alpha_d|^2)
\end{array}
\right) \,.
\label{minormatrix}
\end{eqnarray}
\end{widetext}
The determinant $D_{ij}$ of $M_{ij}$ is given by
\begin{eqnarray*}
D_{ij}&=&Det[M_{ij}] \nonumber\\
& = & -\frac{1}{d^2}(p_i(1-|\alpha_i|^2)+p_j(1-|\alpha_j|^2))\times(p_1|\alpha_1|^2+\cdots\nonumber\\
&  & +p_d|\alpha_d|^2)\times |p_i\alpha_i\sqrt{1-|\alpha_i|^2}-p_j\alpha_j\sqrt{1-|\alpha_j|^2}|^2.
\end{eqnarray*}
$D_{ij}\geq 0$ implies that
\begin{eqnarray}\label{conditi}
p_1\alpha_1\sqrt{1-|\alpha_1|^2}=\cdots=p_d\alpha_d\sqrt{1-|\alpha_d|^2}.
\end{eqnarray}

In the protocol of assisted state discrimination,
$p_1,p_2,\cdots, p_d$ are the \emph{priori} probabilities. Since  the overlap
$\alpha_{1i}=\langle\psi_1|\psi_i\rangle=\alpha_1^*\alpha_i$ is fixed, without lose
of generality, we can rewrite (\ref{conditi}) as
\begin{eqnarray}\label{conditii}
p_1\alpha_1\sqrt{1-|\alpha_1|^2}& = & p_2\frac{\alpha_{12}}{\alpha_1^*}\sqrt{1-|\frac{\alpha_{12}}{\alpha_1}|^2}=\cdots\nonumber\\
& = & p_d\frac{\alpha_{1d}}{\alpha_1^*}\sqrt{1-|\frac{\alpha_{1d}}{\alpha_1}|^2}.
\end{eqnarray}
There is only one variable $\alpha_1$ in (\ref{conditii}) but $d-1$ equalities.
By using the condition (\ref{conditi}), we successfully write $\rho$ in the following separable form,
\begin{eqnarray}\label{rhosep}
\rho=\rho_1\otimes |0\rangle_a\langle0|+\rho_2\otimes \rho_a,
\end{eqnarray}
where
\begin{eqnarray*}
\rho_1 & = & p_1(1-|\alpha_1|^2)|1\rangle\langle1|+\cdots+p_d(1-|\alpha_d|^2)|d\rangle\langle d|;\nonumber\\
\rho_2 & = & \frac{1}{d}(|1\rangle+\cdots+|d\rangle)(\langle1|+\cdots+\langle d|);\nonumber\\
\rho_a & = & (p_1|\alpha_1|^2+\cdots+p_d|\alpha_d|^2)|1\rangle_a\langle1|\nonumber\\
        &  & + \sqrt{d}p_1\sqrt{1-|\alpha_1|^2}(\alpha_1|1\rangle_a\langle0|+\alpha_1^*|0\rangle_a\langle1|).
\end{eqnarray*}
Thus condition (\ref{conditi}) is actually also the sufficient condition for separability.

From the necessary and sufficient condition of zero discord
proposed by Eq.(\ref{zeorleftdis}) and (\ref{zeorrightdis}),
$\rho$ has vanishing ``right'' discord if and only if $[|0\rangle_a\langle0|, \rho_a]=0$,
which gives rise to $\alpha_1=0$ or $\alpha_1=1$.
From (\ref{conditi}) it means that
the overlap $\langle\psi_i|\psi_j\rangle$ can only be either $0$ or $1$, namely
$|\psi_i\rangle,|\psi_j\rangle$
are either orthogonal or equal, which contradicts with our assumption.
Therefore we conclude that in assisted
state discrimination, the ``right'' quantum discord is always
required. This agrees with the case of two non-orthogonal states.

We consider now the ``left'' quantum discord.
It is easy to find that $\rho_1, \rho_2$ is linearly independent.
Hence $\rho$ has vanishing ``left'' discord if and only if  $[\rho_1, \rho_2]=0$.
With this commuting condition, i.e., assuming that the ``left'' discord is zero,
we should have
\begin{eqnarray}\label{con12}
p_1(1-|\alpha_1|^2)=\cdots=p_d(1-|\alpha_d|^2).
\end{eqnarray}
It means that $\rho _1$ is an identity and commutes with any density operators.
Combining Eq. (\ref{conditi}) and Eq. (\ref{con12}), we obtain
\begin{eqnarray}
& &  p_1=\cdots=p_d=\frac{1}{d}, \label{cond1} \\
&&
\alpha_1=\cdots=\alpha_d\equiv \gamma .\label{cond2}
%& &  |\alpha_1|=\sqrt{\langle\psi_i|\psi_j\rangle}, ~~i\neq j,
\end{eqnarray}
On the other hand, from the unitary transformation (\ref{dunitary}),
the \emph{priori} fixed overlaps are equal and
should be a real number,
\begin{eqnarray}\label{overlap}
\langle\psi_i|\psi_j\rangle=\alpha_i^*\alpha_j=|\gamma |^2.
\end{eqnarray}

To conclude, the ``right'' discord is always required for the assisted state discrimination of $d$ ($d\geq2$) nonorthogonal states, though the quantum entanglement could be absent, i.e.
the condition (\ref{conditi}) is fulfilled. This is a generalization of assisted state
discrimination for two nonorthogonal states \cite{Roa}.
On the other hand, the ``left'' discord is not necessarily required in this process since conditions (\ref{cond1}) and (\ref{cond2})
can be fulfilled for some cases which lead also to the absence of entanglement.

As a byproduct, one may notice that the PPT condition is a sufficient condition of separability
for state Eq.(\ref{rhod}) since it can be written into the form of (\ref{rhosep}).
On the other hand, PPT is a necessary condition for separability.  Thus PPT criterion is both the necessary and the
sufficient condition for
the separability of a class of $2\times d$ states of the form (\ref{rhod}).

\section{Optimal unambiguous discrimination of  $d$ nonorthogonal states}\label{optimal}
Next, we will try to determine the exact form of the optimal success probability given by (\ref{proo}).
As $p_1,p_2,\cdots, p_d$ are the \emph{priori} probabilities and the overlap
$\alpha_{1i}=\langle\psi_1|\psi_i\rangle=\alpha_1^*\alpha_i$ is known,
the success probability in (~\ref{proo}) can be rewritten  as
\begin{eqnarray}\label{prooo}
P=1-p_1|\alpha_1|^2-\frac{p_2|\alpha_{12}|^2+\cdots+p_d|\alpha_{1d}|^2}{|\alpha_{1}|^2}.
\end{eqnarray}
There is only one variable $|\alpha _1|$ in (\ref{prooo}).  In order to find the optimal probability,
we define
\begin{eqnarray}
\bar{\alpha}=\sqrt[4]{\frac{p_2|\alpha_{12}|^2+\cdots+p_d|\alpha_{1d}|^2}{p_1}}.
\end{eqnarray}
The optimal probability can be found by dealing with $\bar {\alpha }$ in three different
regions:
\begin{enumerate}
\item
If $\max\{|\alpha_{12}|,\cdots,|\alpha_{1d}|\}\leq\bar{\alpha}\leq 1$,
then when $|\alpha_{1}|=\bar{\alpha}$, we have the optimal probability
\begin{eqnarray*}
P=1-2\sqrt{p_1}\sqrt{p_2|\alpha_{12}|^2+\cdots+p_d|\alpha_{1d}|^2}.
\end{eqnarray*}

\item
If $\bar{\alpha}\leq\max\{|\alpha_{12}|,\cdots,|\alpha_{1d}|\}$,
then when  $|\alpha_{1}|=\max\{|\alpha_{12}|,\cdots,|\alpha_{1d}|\}$, $P$ reaches
its optimal point,
\begin{eqnarray*}
P& = & 1-p_1\max\{|\alpha_{12}|^2,\cdots,|\alpha_{1d}|^2\}-\nonumber\\
&  & \frac{p_2|\alpha_{12}|^2+\cdots+p_d|\alpha_{1d}|^2}{\max\{|\alpha_{12}|^2,\cdots,|\alpha_{1d}|^2\}}.
\end{eqnarray*}

\item
If $1\leq\bar{\alpha}$, when $|\alpha_{1}|=1$, we can obtain the optimal probability
\begin{eqnarray*}
P=1-p_1-p_2|\alpha_{12}|^2-\cdots-p_d|\alpha_{1d}|^2.
\end{eqnarray*}

\end{enumerate}

As we have found that the ``right'' discord is always required for assisted state
discrimination, but ``left'' discord
can be zero. In the
following, we will show that the  ``left'' discord vanishes in the optimal process.

As we already show that (\ref{cond1}) and (\ref{cond2}) are the
conditions for vanishing ``left'' discord, in assisted state discrimination,
a qudit is randomly prepared
in one of the $d$ nonorthogonal states $|\psi_i\rangle$ ($i=1,\cdots,d$) with an equal \emph{priori} probability $p_j$
and an equal \emph{priori} and nonzero overlap $\langle\psi_i|\psi_j\rangle =|\gamma |^2$ for all $i\neq j$. Since $|\psi_i\rangle$
is selected in $d$ dimension Hilbert space, this can always be achieved. In this case, the parameter $\bar{\alpha}=\sqrt[4]{d-1}|\gamma |$,
and we also have,
$\max\{|\alpha_{12}|,\cdots,|\alpha_{1d}|\}=|\gamma|^2\leq\bar{\alpha}$, thus
the optimal probability can be expressed as
\begin{equation}\label{euqalopt}
P_{opt}=\left\{\begin{array}{l}
1-\frac{2\sqrt{d-1}}{d}|\gamma |^2,~~~~ 0\leq|\gamma |\leq \frac{1}{\sqrt[4]{d-1}},\\
 \frac{d-1}{d}(1-|\gamma |^4),~~~~ \frac{1}{\sqrt[4]{d-1}} \leq|\gamma |\leq1.
\end{array}\right.
\end{equation}
We see that the assisted optimal state discrimination can be accomplished with a vanishing ``left''
discord. Also from (\ref{euqalopt}),  we have that for any $d$, the optimal probability is
monotonically decreasing with the non-negative parameter $|\gamma|$,
i.e., the smaller the overlap is, the larger probability we can discriminate. This is understandable,  when they
are orthogonal, i.e. the overlap is zero, we can discriminate them deterministically;
when they are close to each other, it is difficult to discriminate them.
In short, we have shown that the optimal assisted state discrimination of $d$ ($d\geq2$) nonorthogonal states can
be performed with the absence of entanglement  and ``left'' discord,  while  the ``right'' discord
is always required.

\medskip
\section{Summary and discussions}\label{discuss}
It is recently known that besides quantum entanglement, other quantum correlations such as the quantum
discord are also useful in quantum information processing. The
physical or operational interpretations of quantum discord are still
under exploration from different points of view , see e.g. \cite{SKB,PWinter}. On the other hand,
in the assisted state discrimination \cite{Roa}, it is known that the quantum discord is
required. Further in this paper, we find that only the ``right'' quantum discord is necessary,
while the ``left'' quantum discord can be zero. This clarifies the role of quantum discord
in assisted state discrimination. In particular, we find that as a resource for quantum information
processing, the use of quantum discord depends on the specified processing task. Explicitly in the process
of assisted state discrimination, the measurement on the right hand side is performed, so
the ``right'' quantum discord is necessary. In the absence of entanglement, if one side quantum discord is zero,
depending on this specified one side measurement, the quantum correlation of discord is like
a classical state. If applied in the assisted state discrimination, it means that the process is
like a classical type. Thus the ``right'' quantum discord is really a useful resource in
this process.

\bigskip
\noindent {\bf Acknowledgments}
This work is supported by NSFC, ``973'' program (2010CB922904) and PHR201007107.

\end{document}